\documentclass[a4paper]{article}
\usepackage[dvips]{graphicx}
\begin{document}

\noindent
{SAGA-HE-175-01}

\bigskip

\centerline{\bf Effective hadron masses and couplings in nuclear matter}
\centerline{\bf and incompressibility}

\bigskip

\centerline{$^*$Kunito Tuchitani$^{1}$, $^{**}$Hiroaki Kouno$^{1}$, Akira Hasegawa$^{1}$ and Masahiro Nakano$^2$}

\noindent
$^1${\it Department of Physics, Saga University, Saga 840, Japan}

\noindent
$^2${\it University of Occupational and Environmental Health, Kitakyusyu 807, Japan}

\noindent
* e-mail address: 00td17@edu.cc.saga-u.ac.jp

\noindent
** e-mail address: kounoh@cc.saga-u.ac.jp

\bigskip

\centerline{\bf Abstract}
The role of effective hadron masses and effective couplings in nuclear matter is studied using a generalized effective Lagrangian for $\sigma$-$\omega$ model. 
A simple relation among the effective masses, the effective couplings and the incompressibility $K$ is derived. 
Using the relation, it is found 
that the effective repulsive and the effective attractive forces are almost canceled to each other at the normal density. 
Inversely, if this cancellation is almost complete, $K$ should be 250$\sim$350MeV. 

\bigskip

%\begin{flushleft}

%%%%%%%%%%%%%%%%%%%%%%%%%%%%%%
\section{Introduction}
%%%%%%%%%%%%%%%%%%%%%%%%%%%%%%

The incompressibility $K$ is a very important macroscopic quantity which characterizes the equation of state (EOS) of nuclear matter. 
In the relativistic $\sigma$-$\omega$ models \cite{rf:Walecka,rf:Chin,rf:Boguta}, it is well-known that the effective nucleon mass $m^*$ is also important for EOS of nuclear matter. \cite{rf:Waldhauser} 
It is also known that $K$ can be related with $m^*$ at the normal density $\rho_0$ as the following relation; \cite{rf:Matsui,rf:Kouno1}
%%%%%%%%
\begin{equation}
K=9\rho_0\left({k_{\rm F}^2\over{3\rho E_{\rm F}^*}}+{g_\omega^2\over{m_\omega^2}}+{m^*\over{E_{\rm F}^*}}{dm^*\over{d\rho}}\right)_{\rho =\rho_0},  
\label{eq:b1}
\end{equation}
%%%%%%%%
where $\rho$, $k_{\rm F}$, $m_\omega$ and $g_\omega$ are the baryon density, the Fermi momentum, the $\omega$-meson mass and the $\omega$-nucleon coupling, respectively, and $E_{\rm F}^*=\sqrt{k_{\rm F}^2+{m^*}^2}$. 
Equation (\ref{eq:b1}) can be expressed by the Landau parameters \cite{rf:Matsui} and can be generalized in the model with the $\omega$-meson self-interaction. \cite{rf:Bodmer,rf:Kouno2} 

Recently, Iwasaki et al. \cite{rf:Iwasaki} have rewritten Eq. (\ref{eq:b1}) as 
%%%%%%%%
\begin{equation}
K=9\rho_0\left({k_{\rm F}^2\over{3\rho E_{\rm F}^*}}+{g_\omega^2\over{m_\omega^2}}-{g_\sigma^2{m^*}^2\over{{{m_\sigma^*}^2E_{\rm F}^*}^2}}\right)_{\rho =\rho_0}, 
\label{eq:b2}
\end{equation}
%%%%%%%%
where $m_\sigma^*$ is an effective $\sigma$-meson mass which is defined with vanishing external momentum. 
Equation (\ref{eq:b2}) is a relation between the macroscopic quantity $K$ and the microscopic quantity $m^*$ and $m_\sigma^*$. 
However, it is somewhat curious that the $\sigma$-nucleon coupling $g_{\rm \sigma}$, the $\omega$-nucleon coupling $g_{\rm \omega}$ and the $\omega$-meson mass $m_\omega$ are constant, while $m^*$ and $m_\sigma^*$ depend on the density. 
In this paper, we generalize Eq. (\ref{eq:b2}) , taking the density dependence of effective couplings and effective $\omega$-meson mass into account. 

This paper is organized as follows.  
In Sec. 2, we generalize the mean field calculation for the generalized effective Lagrangian with a limitless number of effective couplings. 
The effective hadron masses and the effective hadron couplings are defined in the framework of the effective action. 
The equation of motions for the meson mean fields and the equations for the meson self-energies with vanishing external momentum are derived. 
In Sec. 3, a simple relation among the incompressibility, the effective hadron masses and the effective hadron couplings is derived. 
It is also shown that the effective repulsive force and the effective attractive force are almost canceled to each other at the normal density. 
Inversely, if this cancellation is almost complete, $K$ should be 250$\sim$350MeV. 
In Sec. 4, the relation between the equation of states (EOS) of nuclear matter, effective hadron masses and effective hadron couplings is studied in detail. 
Section 5 is devoted to the summary. 

%%%%%%%%%%%%%%%%%%%%%%%%%%%%%%%
\section{Effective Lagrangian and Equation of Motion}
%%%%%%%%%%%%%%%%%%%%%%%%%%%%%%%

We start with the following Lagrangian of $\sigma$-$\omega$ model. 
%%%%%%%%
\begin{eqnarray}
L&=&\bar{\psi}\left[i\gamma^\mu\{\partial_\mu+\Sigma_\mu (\sigma, \omega )\} -\{m+\Sigma_{\rm s}(\sigma ,\omega)\}\right]\psi
\nonumber\\
&+&{1\over{2}}\partial^\mu\sigma\partial_\mu\sigma
-{1\over{4}}F_{\mu\nu}F^{\mu\nu}
-U_M(\sigma,\omega ), 
\label{eq:1}
\end{eqnarray}
%%%%%%%%
where $\psi$, $\sigma$, $\omega^\mu$ and $m$ are the nucleon field, the $\sigma$-meson field, the $\omega$-meson field and the nucleon mass, respectively, and $F_{\mu\nu}=\partial_\mu\omega_\nu -\partial_\nu\omega_\mu$. 
The $\Sigma_{\rm s}$, $\Sigma^\mu$ and $U_{\rm M}$ are functions of $\sigma$ and $\omega^\mu$. 
The $\Sigma_{\rm s}$ and $\Sigma^\mu$ are the meson-nucleon interactions, while the $U_{\rm M}$ is the mesonic potential which includes the meson mass terms, the term of the $\sigma$-meson self-interaction, the term of the $\omega$-meson self-interaction and the term of the $\sigma$-$\omega$ interaction. 

We regard the Lagrangian (\ref{eq:1}) as an effective one in which the quantum effects of the vacuum fluctuations have been already included. 
Therefore, in principle, there are a limitless number of parameters, namely, effective couplings in $\Sigma_{\rm s}$, $\Sigma_\mu$ and $U_{M}$. 

We remark that the effective Lagrangian (\ref{eq:1}) includes the large classes of the relativistic nuclear models. 
It includes the original Walecka model,\cite{rf:Walecka,rf:Matsui} the relativistic Hartree approximation,\cite{rf:Chin} the nonlinear $\sigma$-$\omega$ model with $\sigma$-meson self-interactions \cite{rf:Boguta,rf:Reinhard,rf:Waldhauser,rf:Sharma,rf:Kouno1,rf:Iwasaki} and the $\omega$-meson self-interaction, \cite{rf:Bodmer,rf:Sugahara,rf:Kouno2} the model including $\sigma$-$\omega$ meson interaction \cite{rf:Moncada}, and so forth. 
It also includes the Zimanyi and Mozkowski (ZM) model after the fermion wave function is rescaled. \cite{rf:Zimanyi} 

Starting from the effective Lagrangian (\ref{eq:1}), we calculate the density effects in nuclear matter. 
We use the mean field approximation. 
In the uniform nuclear matter, the ground-state expectation value of the spatial component of the $\omega$-meson field is zero. 
Therefore, below, we only work with the expectation value $<\omega^0>$ of the time component of the $\omega$-meson field and write it in the symbol of $\omega$. 
We also write the expectation value $<\sigma >$ in the symbol of $\sigma$.  

The $\Sigma_{\rm s}$ and $\gamma^\mu\Sigma_\mu$ are the self-energies of the nucleon. (See Fig. 1(a).) 
Since $\Sigma^i(i=1,2,3)$ has at least one spatial component of the $\omega$-field, it also becomes zero in the mean field approximation. 
Below, we write $\Sigma^0$ as $\Sigma_{\rm v}$. 
In the Lagrangian (\ref{eq:1}), we have neglected the other parts of the self-energies which vanish in the mean-field approximation. 
(For example, the tensor part $\bar{\psi}[\gamma^\mu,\gamma^\nu ]\Sigma_{\mu\nu}\psi$ vanishes in the mean field approximation, since $\Sigma_{\mu\nu}$ is antisymmetric in the subscripts $\mu$ and $\nu$ and includes at least one $\omega^i$. )

In the mean field approximation, the nucleon propagator is given as \cite{rf:Walecka, rf:Chin}
%%%%%%%%
\begin{equation}
G(k)=G_{\rm F}(k)+G_{\rm D}(k)
\label{eq:1a}
\end{equation}
%%%%%%%%
with the Feynman part 
%%%%%%%%
\begin{equation}
G_{\rm F}(k)=(\gamma^\mu k_\mu^*+m^*){-1\over{-{k^*}^2+{m^*}^2-i\epsilon}}
\label{eq:1af}
\end{equation}
%%%%%%%%
and the density part 
%%%%%%%%
\begin{equation}
G_{\rm D}(k)=(\gamma^\mu k_\mu^*+m^*){i\pi\over{E_k^*}}\delta ({k^*}^0-E_k^*)\theta (k_{\rm F}-\vert {\bf k}\vert )
\label{eq:1ad}
\end{equation}
%%%%%%%%
where $m^*=m+\Sigma_{\rm s}$, ${k^*}^\mu=(k^0+\Sigma_{\rm v},{\bf k})$ and $E_k^*=\sqrt{{\bf k}^2+{m^*}^2}$, respectively, and $k_{\rm F}$ is the Fermi momentum. 
Since the effects of vacuum fluctuations have been already included in the effective Lagrangian (\ref{eq:1}), we use only the density part $G_{\rm D}(k)$ to evaluate the density effects. 

Using the propagator $G_{\rm D}(k)$, we get 
the baryon density 
\begin{eqnarray}
\rho 
&=& <\bar{\psi}\gamma^0\psi >
=-i\int{d^4k\over{(2\pi)^4}}{\rm Tr}[\gamma^0G_{\rm D}(k)]
={\lambda\over{3\pi^2}}k_{\rm F}^3,  
\label{eq:3}
\end{eqnarray}
where $\lambda =2$ in symmetric nuclear matter. 
The scalar density is also given by 
\begin{eqnarray}
\rho_{\rm s}&=&<\bar{\psi}\psi >
=-i\int{d^4k\over{(2\pi)^4}}{\rm Tr}[G_{\rm D}(k)]
\nonumber\\
&=&{\lambda\over{2\pi^2}}m^*\left\{k_{\rm F}E_{\rm F}^*-{m^*}^2\ln{\left({k_{\rm F}+E_{\rm F}^*\over{m^*}}\right)}\right\}, 
\label{eq:4}
\end{eqnarray}
where $E_{\rm F}^*=\sqrt{k_{\rm F}^2+{m^*}^2}$. 
The energy density of the nuclear matter is given by 
\begin{equation}
\epsilon
=\epsilon_{\rm N}(\rho,m^*)+U_{\rm M}(\sigma,\omega )-\Sigma_{\rm v}(\sigma,\omega )\rho, 
\label{eq:5}
\end{equation}
where 
\begin{eqnarray}
\epsilon_N
&=&{\lambda\over{12\pi^2}}\left\{ E_{\rm F}^* k_{\rm F}(3k_{\rm F}^2+{3\over{2}}{m^*}^2)-{3\over{2}}{m^*}^4\log{\left({E_{\rm F}^*+k_{\rm F}\over{m^*}}\right)}\right\}. 
\nonumber\\
&&\label{eq:6}
\end{eqnarray}
The pressure of the nuclear matter is also given by 
\begin{equation}
P=P_{\rm N}(\rho,m^*)-U_{\rm M}(\sigma,\omega ), 
\label{eq:7}
\end{equation}
where 
\begin{eqnarray}
P_N(\rho,m^*)&=&{\lambda\over{12\pi^2}}\left\{ E_{\rm F}^* k_{\rm F}(k_{\rm F}^2-{3\over{2}}{m^*}^2)+{3\over{2}}{m^*}^4\log{\left({E_{\rm F}^*+k_{\rm F}\over{m^*}}\right)}\right\}. 
\nonumber\\
&&\label{eq:8}
\end{eqnarray}
From the thermodynamical identity $(\epsilon +P)/\rho =\mu$, 
the baryonic chemical potential $\mu$ is given by 
\begin{equation}
\mu =E_{\rm F}^*-\Sigma_{\rm v}(\sigma,\omega). 
\label{eq:9}
\end{equation}

The equation of motion for $\sigma$-meson is given by 
\begin{equation}
{\partial \epsilon (\rho,\sigma,\omega )\over{\partial \sigma}}=0. 
\label{eq:10}
\end{equation}
Putting (\ref{eq:5}) into (\ref{eq:10}), we get 
\begin{eqnarray}
&&{\partial \epsilon_{\rm N}(\rho, m^*(\sigma,\omega))\over{\partial \sigma}}
-{\partial \Sigma_{\rm v}(\sigma,\omega)\over{\partial \sigma}}\rho
+{\partial U_{\rm M}(\sigma,\omega )\over{\partial \sigma}}
\nonumber\\
&=&
{\partial m^*(\sigma,\omega )\over{\partial \sigma}}
{\partial \epsilon_{\rm N}(\rho,m^*)\over{\partial m^*}}
-{\partial \Sigma_{\rm v}(\sigma,\omega)\over{\partial \sigma}}\rho
+{\partial U_{\rm M}(\sigma,\omega )\over{\partial \sigma}}
\nonumber\\
&=&{\partial \Sigma_{\rm s}(\sigma,\omega )\over{\partial \sigma}}\rho_{\rm s}
-{\partial \Sigma_{\rm v}(\sigma,\omega)\over{\partial \sigma}}\rho
+{\partial U_{\rm M}(\sigma,\omega )\over{\partial \sigma}}
\nonumber\\
&=&-g_{\rm s\sigma}^*\rho_{\rm s}+g_{\rm v\sigma}^*\rho 
+{\partial U_{\rm M}(\sigma,\omega )\over{\partial \sigma}}=0, 
\label{eq:11}
\end{eqnarray}
where we have used the relation 
%%%%%%%%
\begin{equation}
{\partial \epsilon_{\rm N}(\rho,m^*)\over{\partial m^*}}=\rho_{\rm s}
\label{eq:11a}
\end{equation}
%%%%%%%%
and have defined the effective couplings for the three-point meson-nucleon interaction as 
%%%%%%%%%
\begin{equation}
g_{\rm s\sigma}^*\equiv 
-{\partial \Sigma_{\rm s}(\sigma,\omega )\over{\partial \sigma}}
~~~~~~~~~{\rm and}~~~~~~~~~
g_{\rm v\sigma}^*\equiv 
-{\partial \Sigma_{\rm v}(\sigma,\omega )\over{\partial \sigma}}. 
\label{eq:13}
\end{equation}
%%%%%%%%%
Note that the differentiating the nucleon self-energies with respect to the meson-field expectation value yields the effective couplings of the meson-nucleon interaction. 
One external line of the meson can be attached at the point where one meson mean field have been removed by the differentiation. 
(See Fig. 1.) 
In general, the effective action is a generating functional of one-particle-irreducible correlation functions, namely, effective masses and effective couplings. \cite{rf:Peskin} 

The equation of motion for the $\omega$-field is given by 
\begin{equation}
{\partial \epsilon (\rho,\sigma,\omega )\over{\partial \omega}}=0. 
\label{eq:14}
\end{equation}
Putting (\ref{eq:5}) into (\ref{eq:14}), we get 
\begin{eqnarray}
&&{\partial \epsilon_{\rm N}(\rho, m^*(\sigma,\omega))\over{\partial \omega}}
-{\partial \Sigma_{\rm v}(\sigma,\omega)\over{\partial \omega}}\rho
+{\partial U_{\rm M}(\sigma,\omega )\over{\partial \omega}}
\nonumber\\
&=&
{\partial m^*(\sigma,\omega )\over{\partial \omega}}
{\partial \epsilon_{\rm N}(\rho,m^*)\over{\partial m^*}}
-{\partial \Sigma_{\rm v}(\sigma,\omega)\over{\partial \omega}}\rho
+{\partial U_{\rm M}(\sigma,\omega )\over{\partial \omega}}
\nonumber\\
&=&{\partial \Sigma_{\rm s}(\sigma,\omega )\over{\partial \omega}}\rho_{\rm s}
-{\partial \Sigma_{\rm v}(\sigma,\omega)\over{\partial \omega}}\rho
+{\partial U_{\rm M}(\sigma,\omega )\over{\partial \omega}}
\nonumber\\
&=&-g_{\rm s\omega}^*\rho_{\rm s}+g_{\rm v\omega}^*\rho 
+{\partial U_{\rm M}(\sigma,\omega )\over{\partial \omega}}=0, 
\label{eq:15}
\end{eqnarray}
where we have also defined the effective couplings  
%%%%%%%%%%%%%%%%%
\begin{equation}
g_{\rm s\omega}^*\equiv 
-{\partial \Sigma_{\rm s}(\sigma,\omega )\over{\partial \omega}}
~~~~~~~~~{\rm and}~~~~~~~~~
g_{\rm v\omega}^*\equiv 
-{\partial \Sigma_{\rm v}(\sigma,\omega )\over{\partial \omega}}. 
\label{eq:17}
\end{equation}
%%%%%%%%%%%%%%%%%
The diagrammatic description for the fourth lines of Eqs. (\ref{eq:11}) and (\ref{eq:15}) is shown in Fig. 2. 
Although there are a limitless number of parameters in $\Sigma_{\rm s}$ and $\Sigma_{\rm v}$, only four effective couplings appear for the meson-nucleon interaction in Eqs. (\ref{eq:11}) and (\ref{eq:15}). 
If we put $g_{\rm v\sigma}^*=g_{\rm s\omega}^*=0$ and approximate $g_{\rm s\sigma}^*$ and $g_{\rm v\omega}^*$ as constants which are determined at the normal density, we have familiar equations of motion which are used in the original Walecka model, \cite{rf:Walecka,rf:Matsui} the RHA calculation \cite{rf:Chin} and the nonlinear $\sigma$-$\omega$ model. \cite{rf:Boguta,rf:Reinhard,rf:Waldhauser,rf:Bodmer,rf:Sharma,rf:Sugahara,rf:Moncada,rf:Kouno1,rf:Kouno2,rf:Iwasaki}

Since the effective potential is a generating function of one-particle-irreducible correlation functions with vanishing external momentum,\cite{rf:Peskin} 
the second derivatives of the energy density with respect to the meson mean fields yield the effective meson masses. 
From the energy density (\ref{eq:5}), this can be directly shown as follows. 
The propagator $G (k)$ can be rewritten as 
%%%%%%%%
\begin{equation}
G(k)=G_-(k)P_-(k_0,\vert{\bf k}\vert )+G_+(k)P_+(k_0,\vert{\bf k}\vert ), 
\label{eq:1ad1}
\end{equation}
%%%%%%%%
where
%%%%%%%%
\begin{equation}
G_{\pm}(k)=(\gamma^\mu k_\mu^*+m^*){-1\over{-{k^*}^2+{m^*}^2\pm i\epsilon}}, 
\label{eq:1ad2}
\end{equation}
%%%%%%%%
%%%%%%%%
\begin{equation}
P_+(k_0,\vert{\bf k}\vert )=\theta (k_0) \theta (k_{\rm F}-\vert {\bf k}\vert ),
\label{eq:1ad3}
\end{equation}
%%%%%%%%
and 
%%%%%%%%
\begin{equation}
P_-(k_0,\vert{\bf k}\vert )=1-\theta (k_0) \theta (k_{\rm F}-\vert {\bf k}\vert ). 
\label{eq:1ad3a}
\end{equation}
%%%%%%%%
We remark that $P_+$ and $P_-$ are projection operators which satisfy 
%%%%%%%%
\begin{equation}
P_+P_+=P_+,~~~~~,P_-P_-=P_-~~~~~{\rm and}~~~~~P_-P_+=P_+P_-=0. 
\label{eq:1ad4}
\end{equation}
%%%%%%%%
Differentiating $G_{\pm}$ with respect to $\Sigma_{\rm s}$, we get
%%%%%%%%
\begin{equation}
{\partial G_{\pm}\over{\partial \Sigma_{\rm s}}}=(G_\pm)^2. 
\label{eq:1ad5}
\end{equation}
%%%%%%%%
Since $G_-=G_{\rm F}$, we get 
%%%%%%%%
\begin{equation}
{\partial G_{\rm F}\over{\partial \Sigma_{\rm s}}}=(G_{\rm F})^2. 
\label{eq:1ad5a}
\end{equation}
%%%%%%%%
Using Eqs. (\ref{eq:1ad4}) and (\ref{eq:1ad5}), we get
%%%%%%%%
\begin{eqnarray}
{\partial G \over{\partial \Sigma_{\rm s}}}&=&(G_-)^2P_-+(G_+)^2P_+
\nonumber\\
&=&(G_-P_-)^2+(G_+P_+)^2+G_-G_+P_-P_++G_+G_-P_+P_-
\nonumber\\
&=&(G_-P_-+G_+P_+)^2=(G)^2. 
\label{eq:1ad6}
\end{eqnarray}
%%%%%%%%
Combining Eqs. (\ref{eq:1ad5a}) and (\ref{eq:1ad6}), we get
%%%%%%%%
\begin{eqnarray}
&&{\partial G_{\rm D} \over{\partial \Sigma_{\rm s}}}
={\partial G \over{\partial \Sigma_{\rm s}}}
-{\partial G_{\rm F} \over{\partial \Sigma_{\rm s}}}
=(G)^2-(G_{\rm F})^2
=G_{\rm F}G_{\rm D}+G_{\rm D}G_{\rm F}+G_{\rm D}G_{\rm D} 
\nonumber\\
\label{eq:1ad7}
\end{eqnarray}
%%%%%%%%
Similarly, we get \cite{rf:Chin}
%%%%%%%%
\begin{equation}
{\partial G_{\rm F}\over{\partial \Sigma_{\rm v}}}=G_{\rm F}\gamma^0G_{\rm F} 
\label{eq:1ad8}
\end{equation}
%%%%%%%%
and 
%%%%%%%%
\begin{equation}
{\partial G_{\rm D}\over{\partial \Sigma_{\rm v}}}
=G_{\rm F}\gamma^0G_{\rm D}+G_{\rm D}\gamma^0G_{\rm F}+G_{\rm D}\gamma^0G_{\rm D}. 
\label{eq:1ad8a}
\end{equation}
%%%%%%%%
Using Eqs. (\ref{eq:1ad7}) and (\ref{eq:1ad8a}), and 
differentiating the l.h.s. of the equations of motion (\ref{eq:11}) and (\ref{eq:15}) with respect to the meson-field expectation values, we get the equation for the meson self-energies \cite{rf:Walecka,rf:Chin,rf:Matsui} with vanishing external momentum, namely, 
%%%%%%%%
\begin{eqnarray}
{\partial^2 \epsilon (\rho,\sigma,\omega)\over{\partial \sigma^2}}
&=&-i{g_{\rm s\sigma}^*}^2\int{d^4k\over{(2\pi)^4}}
{\rm Tr}[G_{\rm F}(k)G_{\rm D}(k)+G_{\rm D}(k)G_{\rm F}(k)+G_{\rm D}(k)G_{\rm D}(k)]
\nonumber\\
&+&ig_{\rm s\sigma\sigma}^*\int{d^4k\over{(2\pi)^4}}{\rm Tr}[G_{\rm D}(k)]
-ig_{\rm v\sigma\sigma}^*\int{d^4k\over{(2\pi)^4}}{\rm Tr}[\gamma^0G_{\rm D}(k)]\nonumber\\
&+&{\partial^2 U_{\rm M}(\sigma,\omega)\over{\partial \sigma^2}}, 
\label{eq:c1}
\end{eqnarray}
%%%%%%%%%
%%%%%%%%
\begin{eqnarray}
{\partial^2 \epsilon (\rho,\sigma,\omega)\over{\partial \sigma\partial\omega}}
&=&-i{g_{\rm s\sigma}^*}{g_{\rm s\omega}^*}\int{d^4k\over{(2\pi)^4}}
{\rm Tr}[G_{\rm F}(k)G_{\rm D}(k)+G_{\rm D}(k)G_{\rm F}(k)+G_{\rm D}(k)G_{\rm D}(k)]
\nonumber\\
&+&ig_{\rm s\sigma\omega}^*\int{d^4k\over{(2\pi)^4}}{\rm Tr}[G_{\rm D}(k)]
-ig_{\rm v\sigma\omega}^*\int{d^4k\over{(2\pi)^4}}{\rm Tr}[\gamma^0G_{\rm D}(k)]\nonumber\\
&+&{\partial^2 U_{\rm M}(\sigma,\omega)\over{\partial \sigma\partial\omega}} 
\label{eq:c5}
\end{eqnarray}
%%%%%%%%%
and 
%%%%%%%%
\begin{eqnarray}
{\partial^2 \epsilon (\rho,\sigma,\omega)\over{\partial \omega^2}}
&=&-i{g_{\rm s\omega}^*}^2\int{d^4k\over{(2\pi)^4}}
{\rm Tr}[G_{\rm F}(k)G_{\rm D}(k)+G_{\rm D}(k)G_{\rm F}(k)+G_{\rm D}(k)G_{\rm D}(k)]
\nonumber\\
&+&ig_{\rm s\omega\omega}^*\int{d^4k\over{(2\pi)^4}}{\rm Tr}[G_{\rm D}(k)]
-ig_{\rm v\omega\omega}^*\int{d^4k\over{(2\pi)^4}}{\rm Tr}[\gamma^0G_{\rm D}(k)]\nonumber\\
&+&{\partial^2 U_{\rm M}(\sigma,\omega)\over{\partial \omega^2}},  
\label{eq:c2}
\end{eqnarray}
%%%%%%%%%
where we have defined the effective couplings for the four-point meson-nucleon interaction as 
%%%%%%%%%
\begin{eqnarray}
g_{\rm s\sigma\sigma}^*&\equiv& 
-{\partial^2 \Sigma_{\rm s}(\sigma,\omega )\over{\partial \sigma^2}},~~~~~~~~
g_{\rm v\sigma\sigma}^*\equiv 
-{\partial^2 \Sigma_{\rm v}(\sigma,\omega )\over{\partial \sigma^2}},~~~~~~~~
\nonumber\\
g_{\rm s\sigma\omega}^*&\equiv& 
-{\partial^2 \Sigma_{\rm s}(\sigma,\omega )\over{\partial \sigma\partial\omega}},~~~~~~~~
g_{\rm v\sigma\omega}^*\equiv 
-{\partial^2 \Sigma_{\rm v}(\sigma,\omega )\over{\partial \sigma\partial\omega}},~~~~~~~~
\nonumber\\
g_{\rm s\omega\omega}^*&\equiv& 
-{\partial^2 \Sigma_{\rm s}(\sigma,\omega )\over{\partial \omega^2}}
~~~~~~~~{\rm and}~~~~~~~~
g_{\rm v\omega\omega}^*\equiv 
-{\partial^2 \Sigma_{\rm v}(\sigma,\omega )\over{\partial \omega^2}}. 
\label{eq:c4}
\end{eqnarray}
%%%%%%%%%
The diagrammatic descriptions for Eqs. (\ref{eq:c1}), (\ref{eq:c5}) and (\ref{eq:c2}) are shown in  Fig. 3. 
If the mixing part ${\partial^2 \epsilon\over{\partial \sigma\partial\omega}}$ can be neglected, ${\partial^2 \epsilon\over{\partial \sigma^2}}$ and 
${\partial^2 \epsilon\over{\partial \omega^2}}$ are the effective meson masses which are defined at the zero external momentum.

%%%%%%%%%%%%%%%%%%%%%%%%%%%%%%
\section{Incompressibility}
%%%%%%%%%%%%%%%%%%%%%%%%%%%%%%

In this section, we derive the simple relation among the effective masses, effective couplings and the incompressibility. 
The incompressibility $K$ is defined by 
\begin{equation}
K=
9\rho_0^2\left.{\partial^2 (\epsilon /\rho ) \over{\partial \rho^2}}\right|_{\rho =\rho_0}
=
9\left.{\partial P\over{\partial \rho}}\right|_{\rho =\rho_0}
=
9\rho_0\left.{\partial \mu\over{\partial \rho}}\right|_{\rho =\rho_0}. 
\label{eq:18}
\end{equation}
Using 
$\mu =E_{\rm F}^*-\Sigma_{\rm v}=\sqrt{k_{\rm F}^2+{m^*}^2}-\Sigma_{\rm v}$, 
we get 
\begin{eqnarray}
{1\over{\rho}}{dP\over{d\rho}}={d\mu\over{d\rho}}
&=&
{dk_{\rm F}\over{d\rho}}{k_{\rm F}\over{E_{\rm F}^*}}
+{m^*\over{E_{\rm F}^*}}\left(
 {\partial \Sigma_{\rm s}(\sigma,\omega )\over{\partial \sigma}}{d\sigma\over{d\rho}}
+{\partial \Sigma_{\rm s}(\sigma,\omega )\over{\partial \omega}}{d\omega\over{d\rho}}\right)
\nonumber\\
&-&\left(
 {\partial \Sigma_{\rm v}(\sigma,\omega )\over{\partial \sigma}}{d\sigma\over{d\rho}}
+{\partial \Sigma_{\rm v}(\sigma,\omega )\over{\partial \omega}}{d\omega\over{d\rho}}\right)
\nonumber\\
&=&{k_{\rm F}^2\over{3\rho E_{\rm F}^*}}+{m^*\over{E_{\rm F}^*}}
\left(-g_{\rm s\sigma}^*{d\sigma\over{d\rho}}-g_{\rm s\omega}^*{d\omega\over{d\rho}}\right)
\nonumber\\
&+&
\left(g_{\rm v\sigma}^*{d\sigma\over{d\rho}}+g_{\rm v\omega}^*{d\omega\over{d\rho}}\right)
\nonumber\\
&=&{k_{\rm F}^2\over{3\rho E_{\rm F}^*}}+^t{\bf g}^* {\bf \Phi}^\prime, 
\label{eq:19}
\end{eqnarray}
where 
\begin{equation}
{\bf g}^*\equiv -{m^*\over{E_{\rm F}^*}}{\bf g}_{\rm s}^*+{\bf g}_{\rm v}^*, 
\label{eq:20}
\end{equation}
\begin{equation}
{\bf g}_{\rm s}^*\equiv 
\left[
\begin{array}{c}
g_{\rm s\sigma}^* \\
g_{\rm s\omega}^*
\end{array}
\right], 
~~~~~~
{\bf g}_{\rm v}^*\equiv 
\left[
\begin{array}{c}
g_{\rm v\sigma}^* \\
g_{\rm v\omega}^* 
\end{array}
\right]
\label{eq:21}
\end{equation}
and 
\begin{equation}
{\bf \Phi}^\prime 
\equiv 
\left[
\begin{array}{c}
{d\sigma\over{d\rho}} \\
{d\omega\over{d\rho}} 
\end{array}
\right]. 
\label{eq:22}
\end{equation}
We remark that the inverse of the effective gamma factor $\gamma^*=E_{\rm F}^*/m^*$ appears in ${\bf g}^*$. 
That is a relativistic effect. \cite{rf:Matsui} 

To determine ${d\sigma\over{d\rho}}$ and ${d\omega\over{d\rho}}$, 
we differentiate the equations of motion 
(\ref{eq:10}) and (\ref{eq:14}) with respect to $\rho$. 
\begin{equation}
{\partial\over{\partial \rho}}{\partial \epsilon (\rho,\sigma,\omega)\over{\partial \sigma}}+{d\sigma\over{d\rho}}{\partial^2\epsilon (\rho,\sigma,\omega )\over{\partial \sigma^2}}+{d\omega\over{d\rho}}{\partial^2 \epsilon (\rho,\sigma,\omega )\over{\partial \sigma\partial \omega}}=0
\label{eq:23}
\end{equation}
\begin{equation}
{\partial\over{\partial \rho}}{\partial \epsilon (\rho,\sigma,\omega)\over{\partial \omega}}+{d\omega\over{d\rho}}{\partial^2\epsilon (\rho,\sigma,\omega )\over{\partial \omega^2}}+{d\sigma\over{d\rho}}{\partial^2 \epsilon (\rho,\sigma,\omega )\over{\partial \sigma\partial \omega}}=0
\label{eq:24}
\end{equation}

We also get 
\begin{eqnarray}
{\partial\over{\partial \rho}}{\partial \epsilon (\rho,\sigma,\omega )\over{\partial
\sigma }}
&=&
{\partial\over{\partial \rho}}{\partial \epsilon_N (\rho, m^*(\sigma,\omega ))\over{\partial\sigma }}
-
{\partial\over{\partial \rho}}\left({\partial \Sigma_{\rm v}(\sigma,\omega)\over{\partial \sigma}}\rho\right)
\nonumber\\
&=&
{\partial\over{\partial \rho}}\left({\partial \Sigma_{\rm s}(\sigma,\omega)\over{\partial \sigma}}{\partial \epsilon_{\rm N} (\rho, m^*)\over{\partial m^* }}\right)
-
{\partial \Sigma_{\rm v}(\sigma,\omega)\over{\partial \sigma}}
\nonumber\\
&=&
-g_{\rm s\sigma}^*{\partial \rho_{\rm s}(\rho, m^*)\over{\partial \rho}}+g_{\rm v\sigma}^* =-g_{\rm s\sigma}^*{m^*\over{E_{\rm F}^*}}+g_{\rm v\sigma}^*
\nonumber\\
&&\label{eq:25}
\end{eqnarray}
and 
\begin{eqnarray}
{\partial\over{\partial \rho}}{\partial \epsilon (\rho,\sigma,\omega )\over{\partial
\omega }}
&=&
{\partial\over{\partial \rho}}{\partial \epsilon_N (\rho, m^*(\sigma,\omega ))\over{\partial \omega }}
-{\partial\over{\partial \rho}}\left({\partial \Sigma_{\rm v}(\sigma,\omega)\over{\partial \omega}}\rho\right)
\nonumber\\
&=&
{\partial\over{\partial \rho}}\left({\partial \Sigma_{\rm s}(\sigma,\omega)\over{\partial \omega}}{\partial \epsilon_N (\rho, m^*)\over{\partial m^* }}\right)
-
{\partial \Sigma_{\rm v}(\sigma,\omega)\over{\partial \omega}}
\nonumber\\
&=&
-g_{\rm s\omega}^*{\partial \rho_{\rm s}(\rho, m^*)\over{\partial \rho}}+g_{\rm v\omega}^* =-g_{\rm s\omega}^*{m^*\over{E_{\rm F}^*}}+g_{\rm v\sigma}^*, 
\nonumber\\
&&\label{eq:27}
\end{eqnarray}
where we have used the relation 
\begin{equation}
{\partial \rho_{\rm s}(\rho, m^*)\over{\partial \rho}}
={m^*\over{E_{\rm F}^*}}={\gamma^*}^{-1}. 
\label{eq:26}
\end{equation}
Using Eqs. (\ref{eq:25}) and (\ref{eq:27}), 
the equations 
(\ref{eq:23}) and (\ref{eq:24}) can be rewritten as 
\begin{equation}
{M^*}^2{\bf \Phi}^\prime =-{\bf g}^*, 
\label{eq:28}
\end{equation}
where the effective mass matrix 
${M^*}^2$ is defined by 
\begin{equation}
{M^*}^2
=\left[
\begin{array}{cc}
{m_{\sigma}^*}^2 & {m_{\sigma\omega}^*}^2 \\
{m_{\sigma\omega}^*}^2 & -{m_{\omega}^*}^2 
\end{array}
\right]
\equiv 
\left[
\begin{array}{cc}
{\partial^2 \epsilon\over{\partial \sigma^2}} & {\partial^2 \epsilon\over{\partial \sigma \partial\omega}} \\
{\partial^2 \epsilon\over{\partial\sigma\partial\omega}} & {\partial^2 \epsilon\over{\partial \omega^2}}
\end{array}
\right] .
\label{eq:29}
\end{equation}

From Eq. 
(\ref{eq:28}), we get 
\begin{equation}
{\bf \Phi}^\prime =-({M^*}^2)^{-1}{\bf g}^*
\label{eq:30}
\end{equation}
or, more concretely, 
\begin{equation}
{d\sigma\over{d\rho}}={-{m_{\omega}^*}^2{g}^*_1-{m_{\sigma\omega}^*}^2{g}^*_2\over{({m_{\sigma\omega}^*}^2)^2+{m_{\sigma}^*}^2{m_{\omega}^*}^2}}
~~~~~{\rm and}~~~~~
{d\omega\over{d\rho}}={{m_{\sigma}^*}^2{g}^*_2-{m_{\sigma\omega}^*}^2{g}^*_1\over{({m_{\sigma\omega}^*}^2)^2+{m_{\sigma}^*}^2{m_{\omega}^*}^2}},  
\label{eq:31}
\end{equation}
where $g^*_1$ and $g^*_2$ are the first and second components of the effective coupling vector ${\bf g}^*$, respectively. 

Putting Eq. (\ref{eq:30}) into Eq. (\ref{eq:19}), we get 
\begin{equation}
{1\over{\rho}}{dP\over{d\rho}}={d\mu\over{d\rho}}={k_{\rm F}^2\over{3\rho E_{\rm F}^*}}-^t{\bf g}^*({M^*}^2)^{-1}{\bf g}^*. 
\label{eq:32}
\end{equation}
Therefore, we get the relation among the effective masses, the effective couplings and the incompressibility. 
\begin{equation}
K=9\rho_0\left.\left({k_{\rm F}^2\over{3\rho E_{\rm F}^*}}-^t{\bf g}^*({M^*}^2)^{-1}{\bf g}^*\right)\right|_{\rho =\rho_0}. 
\label{eq:33}
\end{equation}

In particular, if there is no effective $\sigma$-$\omega$ interaction, 
\begin{equation}
-{\bf g}^*({M^*}^2)^{-1}{\bf g}^*
={{g_{\rm v\omega}^*}^2\over{{m_{\omega}^*}^2}}-{{g_{\rm s\sigma}^*}^2{m^*}^2\over{{m_{\sigma}^*}^2{E_{\rm F}^*}^2}}
\label{eq:33a}
={{g_{\rm v\omega}^*}^2\over{{m_{\omega}^*}^2}}-{{g_{\rm s\sigma}^*}^2\over{{m_{\sigma}^*}^2}}{\gamma^*}^{-2}. 
\end{equation}
Therefore, this quantity represents the difference between the strengths of the effective repulsive force and the effective attractive force. 
In the original Walecka model, \cite{rf:Walecka} this quantity can be related to the Landau parameter. \cite{rf:Matsui} 

From Eq. (\ref{eq:33}), we get 
\begin{equation}
-\left.^t{\bf g}^*({M^*}^2)^{-1}{\bf g}^*\right|_{\rho =\rho_0}
={K\over{9\rho_0}}-\left.{k_{\rm F}^2\over{3\rho\sqrt{k_{\rm F}^2+{m^*}^2}}}\right|_{\rho =\rho_0}.
\label{eq:34}
\end{equation}
Equation (\ref{eq:34}) means that 
$-\left.^t{\bf g}^*({M^*}^2)^{-1}{\bf g}^*\right|_{\rho =\rho_0}$
 is the function of $\rho_0$, $m^*$ and $K$ only. 
In fig. 4, we show this quantity as a function of $m^*$. 
As $m^*$ increases, $-\left.^t{\bf g}^*({M^*}^2)^{-1}{\bf g}^*\right|_{\rho =\rho_0}$ increases. 
However, in all cases, the value of this quantity is at most of the order of 10GeV$^{-2}$. 

Furthermore, 
if the $\sigma$-$\omega$ interaction can be neglected and there is no higher contribution of $\omega$-field than ones in the original Walecka model, \cite{rf:Walecka,rf:Matsui} 
$g_{\rm v\omega}^*/m_\omega^*$ becomes a constant and is determined by \cite{rf:Boguta,rf:Waldhauser,rf:Kouno1} 
\begin{equation}
{{g_{\rm v\omega}^*}^2\over{{{m_{\omega}}^*}^2}}=\left.{m-a_1-\sqrt{k_{\rm F}^2+{m^*}^2}\over{\rho_0}}\right|_{\rho =\rho_0},   
\label{eq:35}
\end{equation}
where $a_1$ is the binding energy per nucleon at the normal density. 
Using (\ref{eq:34}) and (\ref{eq:35}), we can determine 
$g_{\rm s\sigma}^*/m_\sigma^*$ by 
\begin{equation}
\left.{{g_{\rm s\sigma}^*}^2\over{{m_{\sigma}^*}^2}}\right|_{\rho =\rho_0}
={k_{\rm F}^2+{m^*}^2\over{{m^*}^2}}\left({{g_{\rm v\omega}^*}^2\over{{{m_{\omega}}^*}^2}}-{K\over{9\rho_0}}+\left.{k_{\rm F}^2\over{3\rho\sqrt{k_{\rm F}
^2+{m^*}^2}}}\right)
\right|_{\rho =\rho_0}. 
\label{eq:36}
\end{equation}
In Fig. 5, we show ${g_{\rm s\sigma}^*}^2/{m_\sigma^*}^2$ and 
${g_{\rm v\omega}^*}^2/{m_\omega^*}^2$ as a function of $m^*$. 
Comparing Fig. 5 with Fig. 4, we see that the effective repulsive force and the effective attractive force are almost canceled to each other at the saturation point. In table 1, we summarize the values of $-^t{\bf g}^*({M^*}^2)^{-1}{\bf g}^*$ at the normal density for the well-known parameter sets (P.S.). 
The NL1 \cite{rf:Reinhard} and NL-SH \cite{rf:Sharma} are the P.S. for the $\sigma$-$\omega$ model with the $\sigma$-meson self-interactions, while TM1 \cite{rf:Sugahara} is the P.S. for the the $\sigma$-$\omega$ model which includes not only the the $\sigma$-meson self-interactions but also the $\omega$-meson self-interaction. 
In all case, the effective repulsive force and the effective attractive force are almost canceled to each other. 

%%%%%%%%%%%%%%%%%%%%%%%% table %%%%%%%%%%%%%%%%%%%%%%%%%%%%%%%%%%%%%%%%%%%%
\begin{table}[ht]
\begin{center}
\begin{tabular}{lccccc} \hline \hline
  P.S.   &  ${g_{\rm s\sigma}^*}^2/{m_\sigma^*}^2$ & ${g_{\rm v\omega}^*}^2/{m_\omega^*}^2$ & $-{\bf g}^*({M^*}^2)^{-1}{\bf g}^*$ 
 & $K$(MeV) & $m^*/m$ \\ \hline
NL1      &  358.390 & 278.979  & -12.420  &  211 & 0.573  \\ 
\hline 
NL-SH    &  325.687 & 273.326  &  3.0916  &  355 & 0.597  \\ 
\hline 
TM1      &  266.501 & 223.099  &  -2.4926 &  281 & 0.634  \\ 
\hline 
\hline
\end{tabular}
\end{center}
\caption{$-{\bf g}^*({M^*}^2)^{-1}{\bf g}^*$ at the normal density. 
The ${g_{\rm s\sigma}^*}^2/{m_\sigma^*}^2$, ${g_{\rm v\omega}^*}^2/{m_\omega^*}^2$  and $-{\bf g}^*({M^*}^2)^{-1}{\bf g}^*$ are written in GeV$^{-2}$. 
}
\label{Table 1}
\end{table}
%%%%%%%%%%%%%%%%%%%%%%%%%%%%%%%%%%%%%%%%%%%%%%%%%%%%%%%%%%%%%%%%%%%%%%%%%%%

Inversely, incompressibility $K$ depends on the value of $m^*$ and $-{\bf g}^*({M^*}^2)^{-1}{\bf g}^*$. 
In Fig. 6, we show $K$ as a function of $m^*$ with the fixed value of $-{\bf g}^*({M^*}^2)^{-1}{\bf g}^*$. 
In the figure, it is seen that $K$ becomes smaller as the $m^*$ becomes larger. 
If $-{\bf g}^*({M^*}^2)^{-1}{\bf g}^*=0$ at the normal density, $K=250\sim 350$MeV. 
This result is very interesting, since it is known that $K=250\sim 350$MeV is favorable to account for the compressional properties of the nuclei. \cite{rf:Kouno1,rf:Kouno2,rf:Kouno3} 

%%%%%%%%%%%%%%%%%%%%%%%%%%%%%%%%%%%%%%%%%%%%%%%%%%%%%%%%%%%%%%%%%%%%%%%%%%%%
\section{Effective hadron masses, effective coupling and Equation of states}
%%%%%%%%%%%%%%%%%%%%%%%%%%%%%%%%%%%%%%%%%%%%%%%%%%%%%%%%%%%%%%%%%%%%%%%%%%%%

In this section, we investigate the EOS at general density. 
Below we assume that
%%%%%%%
\begin{equation}
\Sigma_{\rm s}(\sigma ) =-g_\sigma\sigma +g_{\sigma 2}\sigma^2, ~~~~~~~~~\Sigma_{\rm v} (\omega )=-g_\omega\omega +g_{\omega 3}\omega^3
\label{eq:101}
\end{equation}
%%%%%%%
and 
%%%%%%%
\begin{eqnarray}
U_{\rm M}(\sigma ,\omega )&=&{1\over{2}}m_{\sigma}^2\sigma^2+{1\over{3}}c_{\sigma 3}\sigma^3+{1\over{4}}c_{\sigma 4}\sigma^4
-
{1\over{2}}m_{\omega}^2\omega^2-{1\over{4}}c_{\omega 4}\omega^4. 
\label{eq:102}
\end{eqnarray}
%%%%%%%
Under these assumptions, 
the effective meson-nucleon couplings are given by 
%%%%%%%
\begin{equation}
g_{\rm s\sigma}^*=g_{\sigma}-2g_{\sigma 2}\sigma,~~~~~~~g_{\rm v\sigma}^*=0,~~~~~~~,g_{\rm s\omega}^*=0~~~~~~~{\rm and}~~~~~~~g_{\rm v\omega}^*=g_{\omega}-3g_{\omega 3}\omega^2. 
\label{eq:103} 
\end{equation}
%%%%%%%
The effective meson masses are given by  
%%%%%%%%
\begin{eqnarray}
{m^*_\sigma}^2&=&{m_\sigma}^2+2c_{\sigma 3}\sigma+3c_{\sigma 4}\sigma^2
+2g_{\sigma 2}\rho_{\rm s}
\nonumber\\
&+&{{g_{\rm s\sigma}^*}^2\lambda\over{2\pi^2}}\left\{k_FE_F^*+2{k_F{m^*}^2\over{{E_F^*}}}-3{m^*}^2\log{\left({k_F+E_F^*\over{m^*}}\right)}\right\}
\label{eq:103a}
\end{eqnarray}
%%%%%%%%
and
%%%%%%%%
\begin{eqnarray}
{m^*_\omega}^2&=&{m_\omega}^2+3c_{\omega 4}\omega^2
+6g_{\omega 3}\rho
\label{eq:103b}
\end{eqnarray}
%%%%%%%%
The $\sigma\sigma$($\omega\omega\omega$)-nucleon interaction yields positive contribution to the effective $\sigma$($\omega$)-meson mass if the coupling $g_{\sigma 2}$($g_{\omega 3}$) is positive. 

We consider the following four cases. 

\noindent
(A) The meson-nucleon couplings are constant with no $\omega$-meson self-interaction. 
Namely, $g_{\sigma 2}=0$, $g_{\omega 3}=0$ and $c_{\omega 4}=0$. 

\noindent
(B) The meson-nucleon couplings are constant with the $\omega$-meson self-interaction. 
Namely, $g_{\sigma 2}=0$ and $g_{\omega 3}=0$.  

\noindent
(C) The $\omega$-nucleon couplings are constant with no $\omega$-meson self-interaction. 
Namely, $g_{\omega 3}=0$ and $c_{\omega 4}=0$. 

\noindent
(D) The $\sigma$-nucleon couplings are constant with no $\omega$-meson self-interaction. 
Namely, $g_{\sigma 2}=0$ and $c_{\omega 4}=0$. 

We put $m=939$MeV, $\rho _0=0.15$fm$^{-3}$ and $a_1=16.0$MeV. 
We also put $c_{\omega 4}/g_\omega^4=0.004$($g_{\sigma 2}/g_\sigma^2=0.27$GeV$^{-1}$, $g_{\omega 3}/g_\omega^3=0.65$GeV$^{-2}$) in the case of  the P.S. B(C,D). 
The remaining parameters are choosen to reproduce $K=300$MeV and $m^*/m=0.8$ at the normal density. 
The parameter sets are summarized in Table 2. 
Using these parameter sets, we have done numerical calculations. 

%%%%%%%%%%%%%%%%%%%%%%%% table %%%%%%%%%%%%%%%%%%%%%%%%%%%%%%%%%%%%%%%%%%%%
\begin{table}[ht]
\begin{center}
\begin{tabular}{lccccc} \hline \hline
            & A & B & C & D &  \\ \hline
${g_{\sigma}}^2/{m_\sigma}^2$  & 219.098 & 217.380 & 223.429 & 207.730 & (GeV$^{-2}$) \\  \hline 
${g_{\omega}}^2/{m_\omega}^2$  & 112.529 & 113.379 & 112.529 & 117.696 & (GeV$^{-2}$) \\  \hline 
${g_{\sigma 2}}/{g_\sigma}^2$  & 0.0 & 0.0 & 0.27 & 0.0 & (GeV$^{-1}$) \\ \hline 
${g_{\omega 3}}/{g_\omega}^3$  & 0.0 & 0.0 & 0.0 & 0.65 & (GeV$^{-2}$) \\  \hline 
${c_{\sigma 3}}/{g_\sigma}^3$  & 1.9400 & 0.86155 & 0.31302 & -5.0507 & (MeV) \\  \hline 
${c_{\sigma 4}}/{g_\sigma}^4$  & 2.7675 & 3.2396 & 1.1470 & 5.7818 &  $\times 10^{-2}$                \\  \hline 
${c_{\omega 4}}/{g_\omega}^4$  & 0.0 & 0.004 & 0.0 & 0.0 &   \\  \hline 
\hline
\end{tabular}
\end{center}
\caption{Summary for the parameter sets}
\label{Table 2}
\end{table}
%%%%%%%%%%%%%%%%%%%%%%%%%%%%%%%%%%%%%%%%%%%%%%%%%%%%%%%%%%%%%%%%%%%%%%%%%%%

In Figs. 7$\sim$10, we show $m^*$, $m_\sigma^*$, $m_\omega^*$, $g_{\rm s\sigma}^*$ and $g_{\rm v\omega}^*$ as functions of the baryon density. 
In all cases, $m^*$ decreases and $m_\sigma^*$ increases as the density increases. 
For the P.S. B and D, $m_\omega^*$ increases as the density increases, while $m_\omega^*$ is constant for the P.S. A and C. 
For the P.S. C(D), $g_{\rm s\sigma}^*$($g_{\rm v\omega}^*$) decreases as density increases. 
The $g_{\rm s\sigma}^*$($g_{\rm v\omega}^*$) is constant for the P.S. A ,B and D(for the P.S. A,B, and C). 

We emphasize that the effective meson masses $m_\sigma^*$ and $m_\omega^*$ are not the on-shell masses which are defined at the poles of the propagators. 
The on-shell masses of mesons can be reduced in medium, even though $m_\sigma^*$ and $m_\omega^*$ increases as the density increases. \cite{rf:Iwasaki,rf:Sakamoto}

In Figs. 11 and 12, we show the ratios 
$(g_{\rm s\sigma}^*/m_\sigma^*)^2$ and $(g_{\rm v\omega}^*/m_\omega^*)^2$ as functions of the baryon density. 
In all cases, $(g_{\rm s\sigma}^*/m_\sigma^*)^2$ decreases as the density increases. 
For the P.S. B and D, the ratio $(g_{\rm v\omega}^*/m_\omega^*)^2$ decreases as  the density increases, while the ratio is constant for the P.S. A and C. 

In Figs. 13, we show $-^t{\bf g}^*({{M^*}^2})^{-1}{\bf g}^*$ as a function of the baryon density. 
For all cases, this quantity becomes nearly zero at the normal density as is seen in this previous section. 
For the P.S. A and C, it increases and approaches to the constant value 
$(g_{\omega}/m_\omega)^2$, since the ratio $(g_{\rm s\sigma}^*/m_\sigma^*)^2$ approaches to zero at high density. 
For the P.S. B and C, this quantity begins to decrease at high density, since 
not only the ratio $(g_{\rm s\sigma}^*/m_\sigma^*)^2$ but also the ratio $(g_{\rm v\omega}^*/m_\omega^*)^2$ decreases at high density.  
 
We define 
%%%%%%%%
\begin{equation}
K(\rho )=9{\partial P\over{\partial \rho}}
=
9\rho{\partial \mu\over{\partial \rho}}=K_1(\rho )+K_2(\rho ), 
\label{eq:104}
\end{equation}
%%%%%%%%
where
%%%%%%%%
\begin{equation}
K_1(\rho) =3{k_{\rm F}^2\over{E_{\rm F}^*}}
\label{eq:105}
\end{equation}
%%%%%%%%
and
%%%%%%%%
\begin{equation}
K_2(\rho) =-9\rho \left(^t{\bf g}^*({{M^*}^2}^{-1}){\bf g}^*\right). 
\label{eq:106}
\end{equation}
%%%%%%%%

In Fig. 14, we show the ratio $K_2(\rho )/K_1 (\rho )$ as a function of the baryon density. 
It is seen that, on the contrary to the case at the normal density, the contribution of $K_2(\rho )$ is the dominant part of $K(\rho )$ at high density, except for the case of the P.S. D, in which the raito $(g_{\rm v\omega}^*/m_\omega^*)^2$ decreases rapidly at high density. 

Finally , in Figs. 15 and 16, we show the binding energy per nucleon and pressure of nuclear matter as functions of the baryon density. 
The P.S. D yields the softest EOS, since it has the smallest $(g_{\rm v\omega}^*/m_\omega^*)^2$ at high density. 

\section{Summary}

We have studied the relation among the effective hadron masses, the effective hadron couplings and incompressibility(or EOS) using the generalized $\sigma$-$\omega$ model. 
The results obtained here are summarized as follows. 

(1) Although the effective Lagrangian includes a limitless number of parameters, the equations for physical quantity at finite density, e.g., equations of motion, can be written in a few effective couplings. 

(2) The simple relation among the effective hadron masses, the effective couplings and the incompressibility is derived. 

(3) It is founded that the strengths of the effective attractive force and the effective repulsive force are almost canceled to each other at the normal density. 

(4) Inversely, if this cancellation is almost complete, $K$ should be 250$\sim$350MeV. 
This value is consistent with the compressional properties of the nuclei. 
\cite{rf:Kouno1,rf:Kouno2,rf:Kouno3} 

(5) Except for the case of the P.S. D, the contribution of $-^t{\bf g}^*({{M^*}^2}^{-1}){\bf g}^*$ to the EOS is dominant at high density, while it is nearly zero at the normal density. 

(6) The P.S. D yields the softest EOS at high density, since the effective mass of the $\omega$-meson increases and the effective $\omega$-nucleon interaction decreases at high density. 

It may be interesting to study the asymetric nuclear matter and the nuetron star
and examine the role of the effective hadron masses and couplings 
 using the parameters sets such as the P.S. C and D. 

\bigskip

\centerline{\bf Acknowledgement}

\bigskip

K.T. and H. K. thank Prof. H. Yabu for useful discussions and suggestions. 
Authors thank Y. Iwasaki, K. Makino and K. Sakamoto for useful discussions.

%%%%%%%%%%%%%%%%%%%%%%%%%%%%%%%%%%%%%%%%%%%%%%%%%%%%%%%%%%%%%%%%%%%%%%%%%%%

\vfill\eject

%%%%%%%%%%%%%%%%%% Fig 1%%%%%%%%%%%%%%%%%%%%%%%%%%%%%%%%%%%%%%%%%%%%%%%%
\begin{figure}
\begin{center}
\begin{minipage}{.80\linewidth}
	\includegraphics[height=40mm,width=100mm]{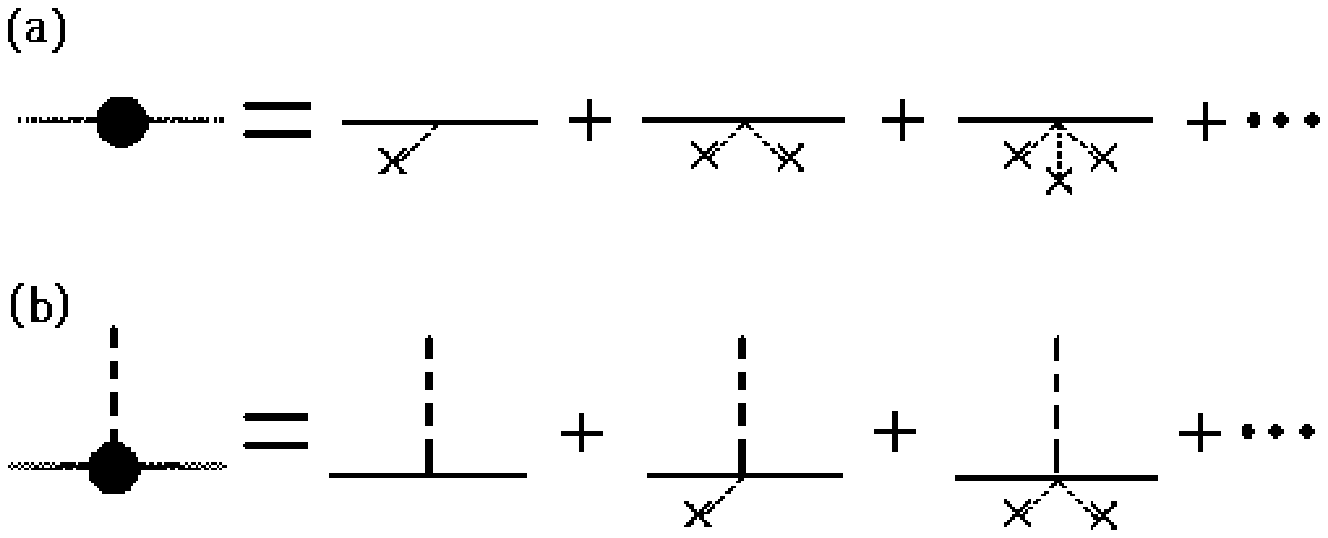}
	\label{fig:01}
\end{minipage}
	\caption{
  The diagramatic descriptions for the nucleon self-energies and 
the effective meson-nucleon couplings. 
The solid and dotted lines represent the nucleon and the meson, respectively. 
The dotted line with the cross represents the meson mean field. 
The solid circle with two nucleon external lines represents the nucleon self-energy, while the solid circle with two nucleon external lines and one meson external line represents the effective meson-nucleon coupling. 
(a) The nucleon self-energies. (b) The effective meson-nucleon couplings. 
}
\end{center}
\end{figure}

%%%%%%%%%%%%%%%%%% Fig 2%%%%%%%%%%%%%%%%%%%%%%%%%%%%%%%%%%%%%%%%%%%%%%%%
\begin{figure}
\begin{center}
\begin{minipage}{.80\linewidth}
	\includegraphics[height=35mm,width=100mm]{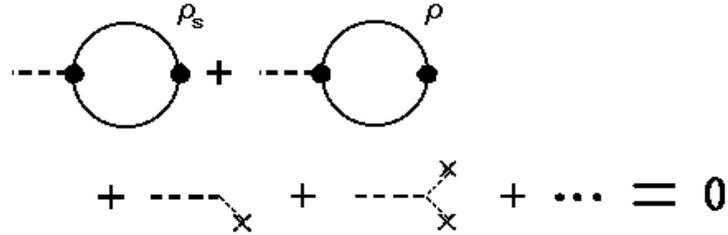}
	\label{fig:02}
\end{minipage}
	\caption{
  The diagramatic description for the equation of motion for mesons. 
The lines,  the cross and the solid circles have the same notation as in Fig. 1.
}
\end{center}
\end{figure}

%%%%%%%%%%%%%%%%%% Fig 3%%%%%%%%%%%%%%%%%%%%%%%%%%%%%%%%%%%%%%%%%%%%%%%%
\begin{figure}
\begin{center}
\begin{minipage}{.80\linewidth}
	\includegraphics[height=35mm,width=100mm]{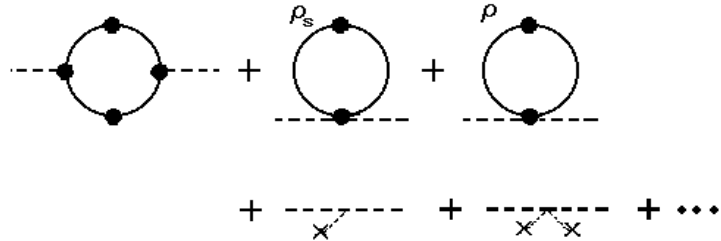}
	\label{fig:03}
\end{minipage}
	\caption{
  The diagramatic description for the meson self-energies. 
The lines, the cross and the solid circles have the same notation as in Fig. 1. 
}
\end{center}
\end{figure}

%%%%%%%%%%%%%%%%%% Fig 4%%%%%%%%%%%%%%%%%%%%%%%%%%%%%%%%%%%%%%%%%%%%%%%%
%\oddsidemargin 5mm
%\evensidemargin 5mm
%\textwidth 190mm

\begin{figure}
\begin{center}
\begin{minipage}{1.20\linewidth}
	\includegraphics[height=75mm,width=95mm]{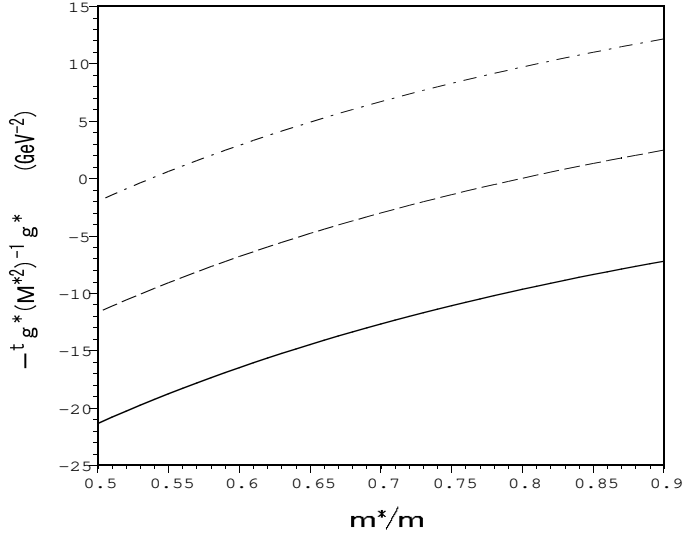}
	\label{fig:04}
	\caption{
  At the normal density, $-^t{\bf g}^*({M^*}^2)^{-1}{\bf g}^*$ is shown as a function of the effective nucleon mass. 
The solid, the dashed and the dash-dotted curves represent the results for $K=$150, 250 and 350MeV, respectively. 
We put $\rho_0=0.15$fm$^{-3}$. 
}
\end{minipage}
\end{center}
\end{figure}

%%%%%%%%%%%%%%%%%% Fig 5%%%%%%%%%%%%%%%%%%%%%%%%%%%%%%%%%%%%%%%%%%%%%%%%
\begin{figure}
\begin{center}
\begin{minipage}{1.20\linewidth}
	\includegraphics[height=75mm,width=95mm]{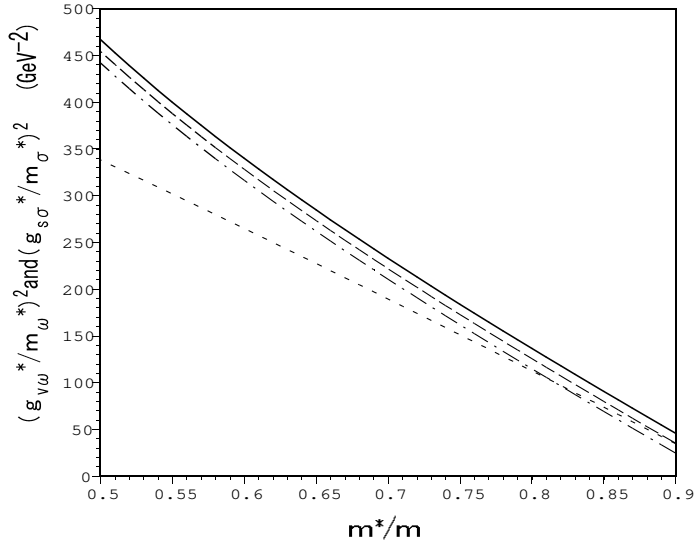}
	\label{fig:05}
	\caption{
  At the normal density, $(g^*_{\rm v\omega}/m_\omega^*)^2$ and $(g^*_{\rm s\sigma}/m_\sigma^*)^2$ are shown as functions of the effective nucleon mass. 
The solid, the dashed and the dash-dotted curves represent the raitos $(g^*_{\rm s\sigma}/m_\sigma^*)^2$ for $K=$150, 250 and 350MeV, respectively. 
The raito $(g^*_{\rm v\omega}/m_\omega^*)^2$ does not depend on the value of $K$ and is represented by the dotted line. 
We put $\rho_0=0.15$fm$^{-3}$ and $a_1=16$MeV. 
}
\end{minipage}
\end{center}
\end{figure}
%%%%%%%%%%%%%%%%%% Fig 6%%%%%%%%%%%%%%%%%%%%%%%%%%%%%%%%%%%%%%%%%%%%%%%%
\begin{figure}
\begin{center}
\begin{minipage}{1.20\linewidth}
	\includegraphics[height=80mm,width=95mm]{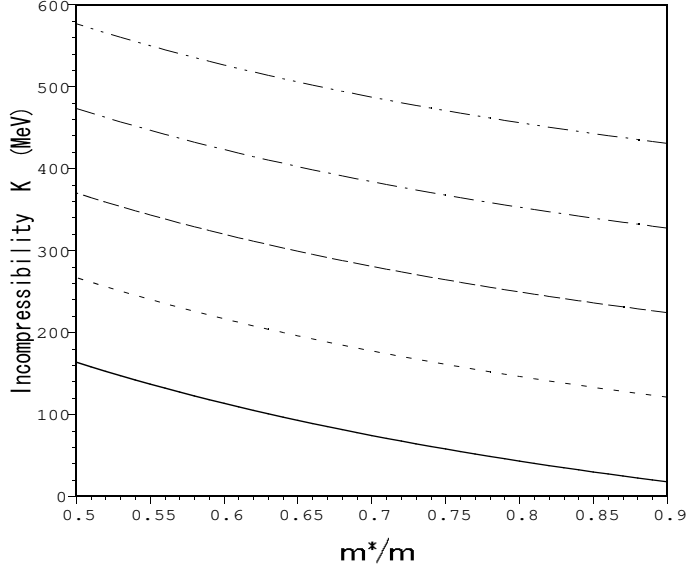}
	\label{fig:06}
	\caption{
  At the normal density, $K$ is shown as a function of the effective nucleon mass. 
The solid, the dotted, the dashed, the dash-dotted and the dash-dot-dotted curves represent the results for $-^t{\bf g}^*({M^*}^2)^{-1}{\bf g}^*=$-20, -10, 0, 10 and 20(GeV$^{-2}$), respectively. 
We put $\rho_0=0.15$fm$^{-3}$. 
}
\end{minipage}
\end{center}
\end{figure}
%%%%%%%%%%%%%%%%%% Fig 7%%%%%%%%%%%%%%%%%%%%%%%%%%%%%%%%%%%%%%%%%%%%%%%%
\begin{figure}
\begin{center}
\begin{minipage}{1.20\linewidth}
	\includegraphics[height=80mm,width=95mm]{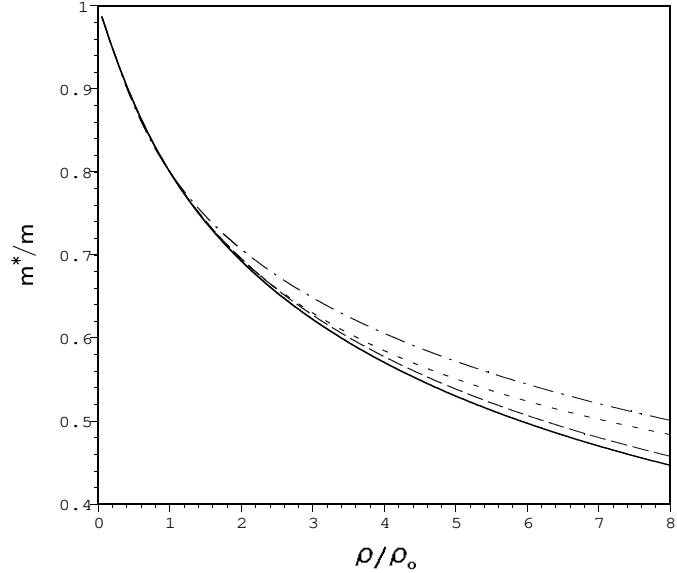}
	\label{fig:07}
	\caption{
  The effective nucleon mass is shown as a function of the baryon density. The solid, the dashed, the dotted and dash-dotted curves represent the results for the P.S. A, B, C and D, respectively. 
}
\end{minipage}
\end{center}
\end{figure}
%%%%%%%%%%%%%%%%%% Fig 8%%%%%%%%%%%%%%%%%%%%%%%%%%%%%%%%%%%%%%%%%%%%%%%%
\begin{figure}
\begin{center}
\begin{minipage}{1.20\linewidth}
	\includegraphics[height=80mm,width=95mm]{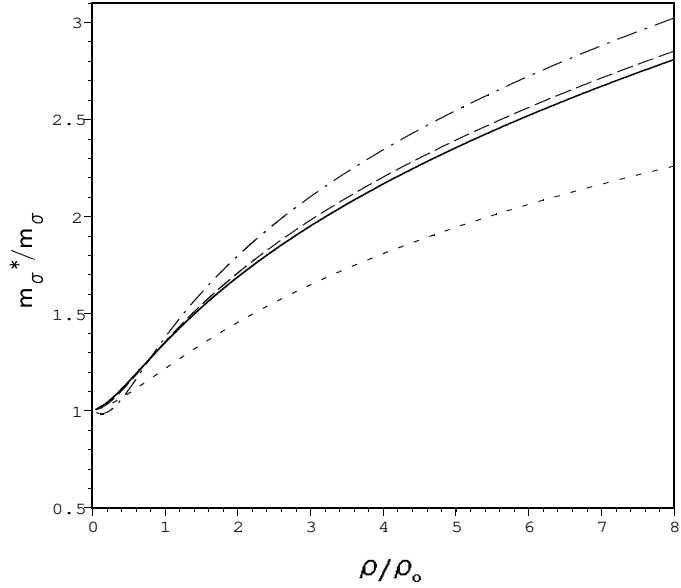}
	\label{fig:08}
	\caption{
  The effective $\sigma$-meson mass is shown as a function of the baryon density. 
The various curves have the same notation as in Fig. 7. 
}
\end{minipage}
\end{center}
\end{figure}
%%%%%%%%%%%%%%%%%% Fig 9%%%%%%%%%%%%%%%%%%%%%%%%%%%%%%%%%%%%%%%%%%%%%%%%
\begin{figure}
\begin{center}
\begin{minipage}{1.20\linewidth}
	\includegraphics[height=80mm,width=95mm]{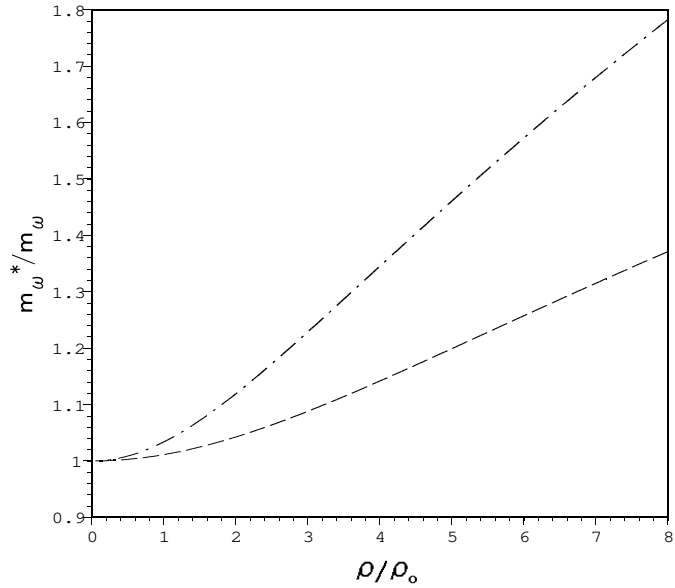}
	\label{fig:09}
	\caption{
  The effective $\omega$-meson mass is shown as a function of the baryon density. 
The dashed and the dash-dotted curves represent the results for the P.S. B and D, respectively. 
}
\end{minipage}
\end{center}
\end{figure}
%%%%%%%%%%%%%%%%%% Fig 10%%%%%%%%%%%%%%%%%%%%%%%%%%%%%%%%%%%%%%%%%%%%%%%%
\begin{figure}
\begin{center}
\begin{minipage}{1.20\linewidth}
	\includegraphics[height=80mm,width=95mm]{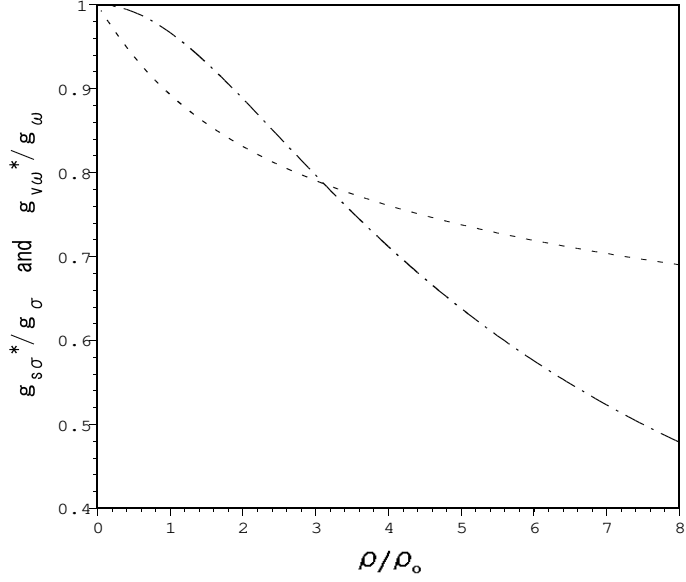}
	\label{fig:10}
	\caption{
  The effective $\sigma$-nucleon coupling and the effective $\omega$-nucleon coupling are shown as functions of the baryon density. 
The dotted and the dash-dotted curves represent the results for $g^*_{\rm s\sigma}$ in the P.S. C and for $g^*_{\rm v\omega}$ in the P.S. D, respectively. 
}
\end{minipage}
\end{center}
\end{figure}
%%%%%%%%%%%%%%%%%% Fig 11%%%%%%%%%%%%%%%%%%%%%%%%%%%%%%%%%%%%%%%%%%%%%%%%
\begin{figure}
\begin{center}
\begin{minipage}{1.20\linewidth}
	\includegraphics[height=80mm,width=95mm]{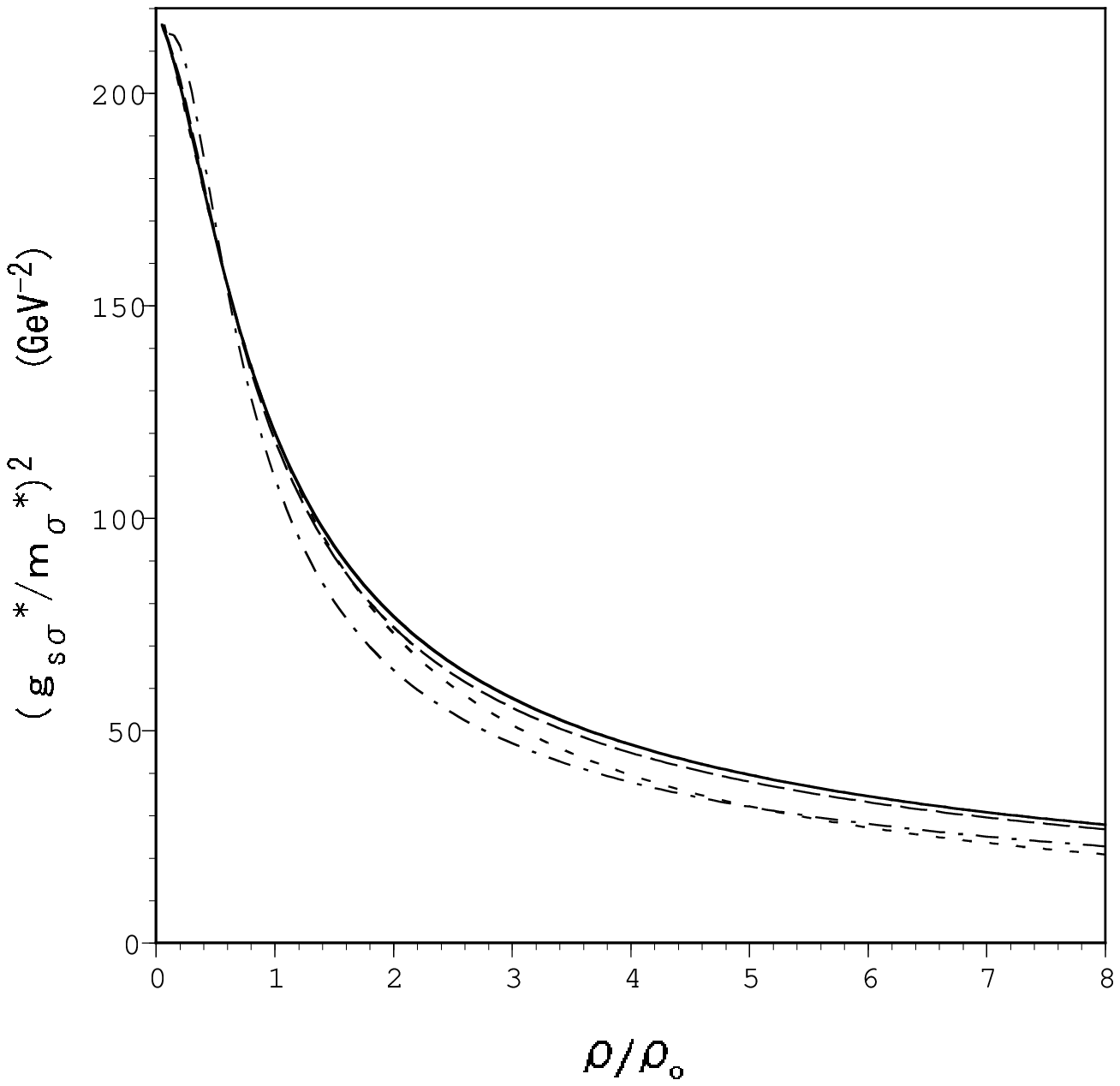}
	\label{fig:11}
	\caption{
  The raito $(g^*_{\rm s\sigma}/m_\sigma^*)^2$ is shown as a function of the baryon density. 
The various curves have the same notation as in Fig. 7. 
}
\end{minipage}
\end{center}
\end{figure}
%%%%%%%%%%%%%%%%%% Fig 12%%%%%%%%%%%%%%%%%%%%%%%%%%%%%%%%%%%%%%%%%%%%%%%%
\begin{figure}
\begin{center}
\begin{minipage}{1.20\linewidth}
	\includegraphics[height=80mm,width=95mm]{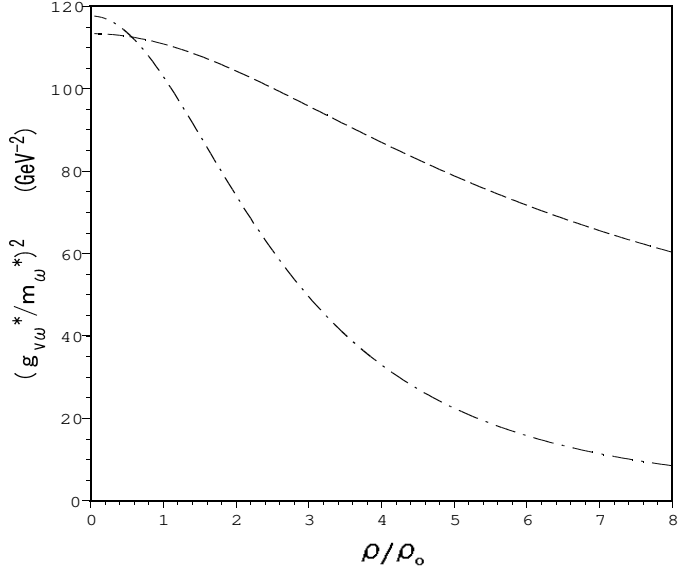}
	\label{fig:12}
	\caption{
  The raito $(g^*_{\rm v\omega}/m_\omega^*)^2$ is shown as a function of the baryon density. 
The dashed and the dash-dotted curves have the same notation as in Fig. 9. 
}
\end{minipage}
\end{center}
\end{figure}
%%%%%%%%%%%%%%%%%% Fig 13%%%%%%%%%%%%%%%%%%%%%%%%%%%%%%%%%%%%%%%%%%%%%%%%
\begin{figure}
\begin{center}
\begin{minipage}{1.20\linewidth}
	\includegraphics[height=80mm,width=95mm]{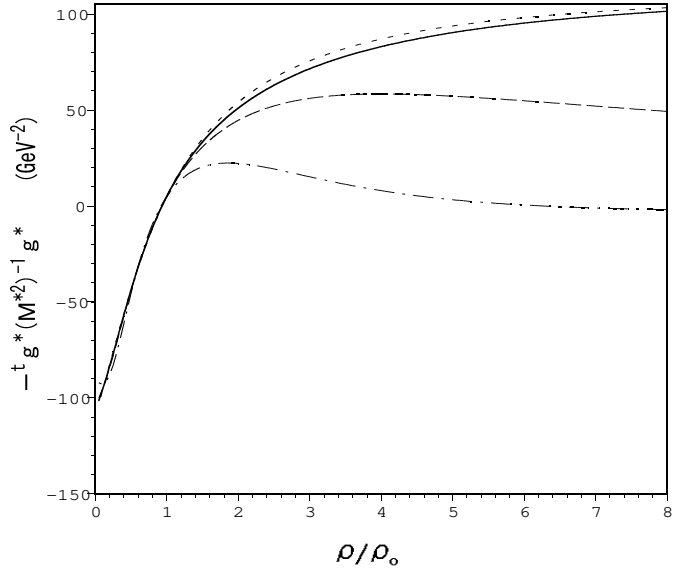}
	\label{fig:13}
	\caption{
  The $-^t{\bf g}^*({M^*}^2)^{-1}{\bf g}^*$ is shown as a function of the baryon density. 
The various curves have the same notation as in Fig. 7. 
}
\end{minipage}
\end{center}
\end{figure}
%%%%%%%%%%%%%%%%%% Fig 14%%%%%%%%%%%%%%%%%%%%%%%%%%%%%%%%%%%%%%%%%%%%%%%%
\begin{figure}
\begin{center}
\begin{minipage}{1.20\linewidth}
	\includegraphics[height=80mm,width=95mm]{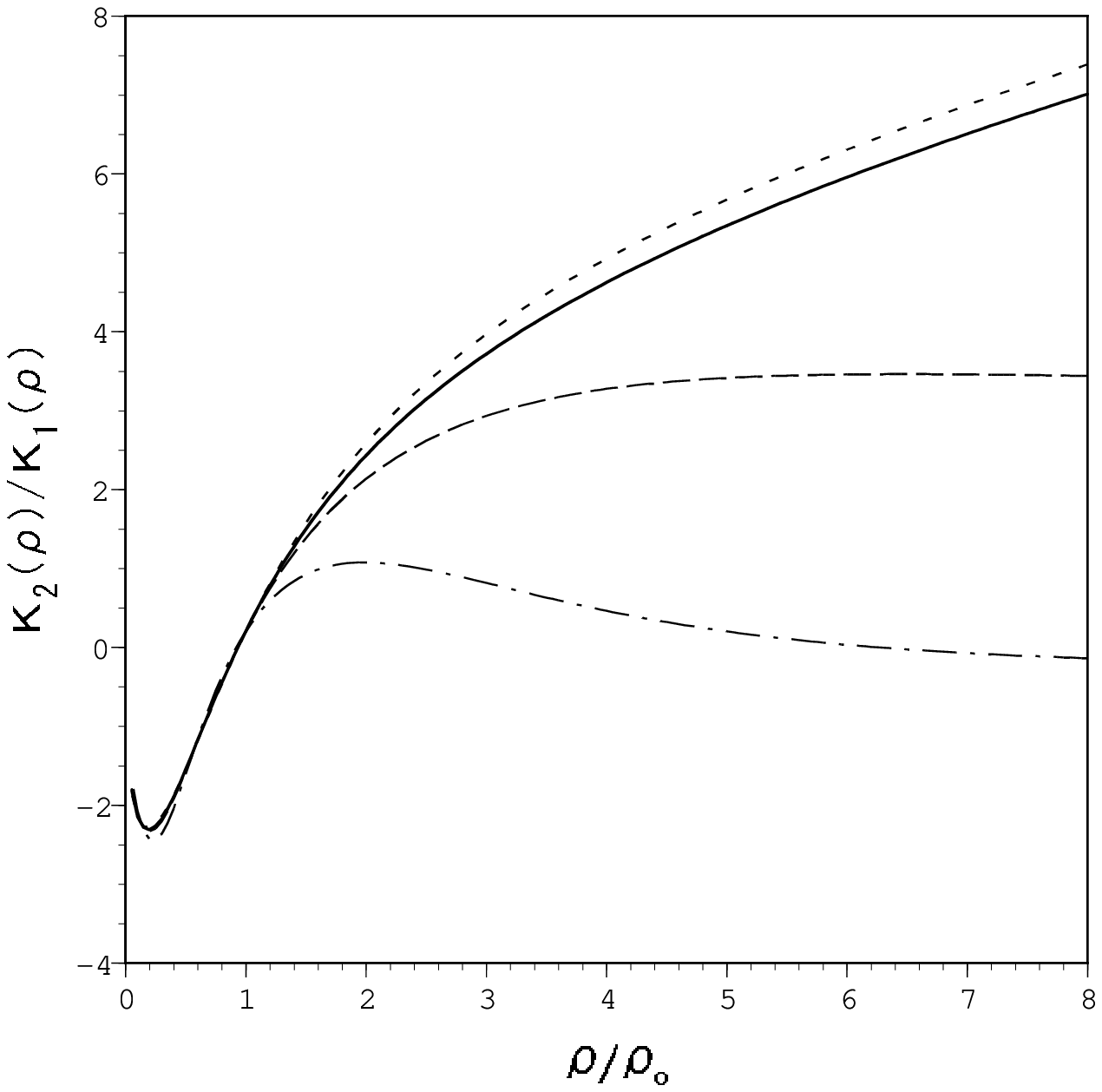}
	\label{fig:14}
	\caption{
  The raito $K_2(\rho)/K_1(\rho)$ is shown as a function of the baryon density. 
The various curves have the same notation as in Fig. 7. 
}
\end{minipage}
\end{center}
\end{figure}
%%%%%%%%%%%%%%%%%% Fig 15%%%%%%%%%%%%%%%%%%%%%%%%%%%%%%%%%%%%%%%%%%%%%%%%
\begin{figure}
\begin{center}
\begin{minipage}{1.20\linewidth}
	\includegraphics[height=80mm,width=95mm]{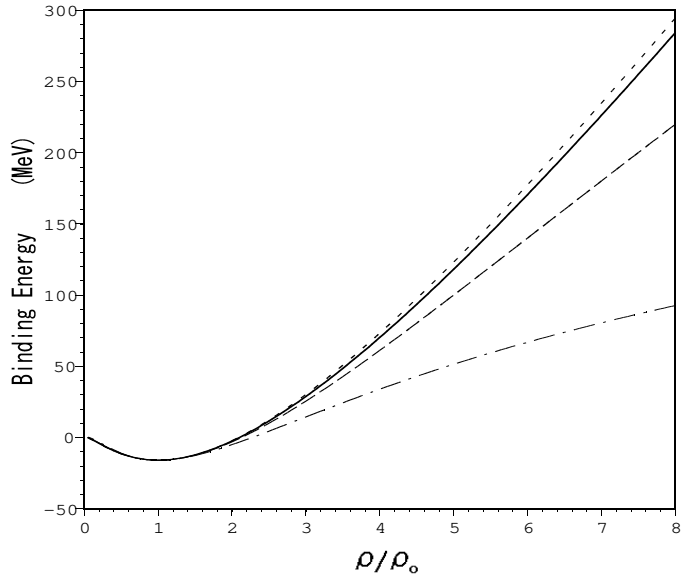}
	\label{fig:15}
	\caption{
  The binding enery per nucleon is shown as a function of the baryon density. 
The various curves have the same notation as in Fig. 7. 
}
\end{minipage}
\end{center}
\end{figure}
%%%%%%%%%%%%%%%%%% Fig 16%%%%%%%%%%%%%%%%%%%%%%%%%%%%%%%%%%%%%%%%%%%%%%%%
\begin{figure}
\begin{center}
\begin{minipage}{1.20\linewidth}
	\includegraphics[height=80mm,width=95mm]{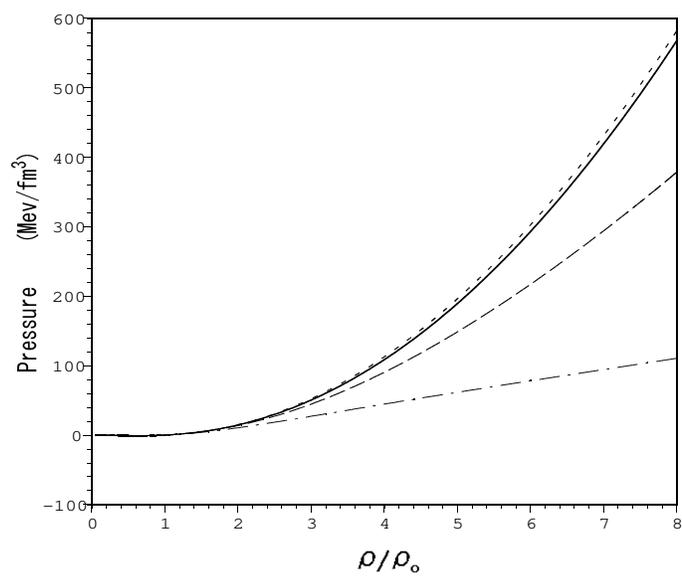}
	\label{fig:16}
	\caption{
  The pressure of nuclear matter is shown as a function of the baryon density. 
The various curves have the same notation as in Fig. 7. 
}
\end{minipage}
\end{center}
\end{figure}

%%%%%%%%%%%%%%%%%%%%%%%%%%%%%%%%%%%%%%%%%%%%%%%%%%%%%%%%%%%%%%%%%%%%%%%%

\end{document}